\documentclass[pra,aps,twocolumn,floatfix,showpacs]{revtex4-1}
\usepackage{graphicx,amsmath,amssymb,times}

\begin{document}
\title{Strong-coupling expansion for the two-species Bose-Hubbard model}
\author{M. Iskin}
\affiliation{Department of Physics, Ko\c c University, Rumelifeneri Yolu, 34450 Sariyer, Istanbul, Turkey}
\date{\today}

\begin{abstract}
To analyze the ground-state phase diagram of Bose-Bose mixtures 
loaded into $d$-dimensional hypercubic optical lattices, we perform 
a strong-coupling power-series expansion in the kinetic energy term 
(plus a scaling analysis) for the two-species Bose-Hubbard model 
with onsite boson-boson interactions. We consider both repulsive 
and attractive interspecies interaction, and obtain an analytical 
expression for the phase boundary between the incompressible 
Mott insulator and the compressible superfluid phase up to third 
order in the hoppings. In particular, we find a re-entrant quantum 
phase transition from paired superfluid (superfluidity of composite 
bosons, i.e. Bose-Bose pairs) to Mott insulator and again to a 
paired superfluid in all one, two and three dimensions, when the 
interspecies interaction is sufficiently large and attractive. We hope 
that some of our results could be tested with ultracold atomic systems.
\end{abstract}

\maketitle

\section{Introduction}
\label{sec:introduction}

Single-species Bose-Hubbard (BH) model is the bosonic generalization 
of the Hubbard model, and was introduced originally to describe $^4$He 
in porous media or disordered granular superconductors~\cite{fisher}. 
For hypercubic lattices in all dimensions $d$, there are only two 
phases in this model: an incompressible Mott insulator at commensurate 
(integer) fillings and a compressible superfluid phase otherwise. 
The superfluid phase is well described by weak-coupling theories, 
but the insulating phase is a strong-coupling phenomenon that only 
appears when the system is on a lattice. Transition from the Mott
insulator to the superfluid phase occurs as the hopping, particle-particle 
interaction, or the chemical potential is varied~\cite{fisher}.

It is the recent observation of this transition in effectively 
three-~\cite{greiner02}, one-~\cite{stoferle04}, and 
two-dimensional~\cite{spielman07, spielman08} optical lattices, 
which has been considered one of the most remarkable 
achievements in the field of ultracold atomic gases, since it
paved the way for studying other strongly correlated phases in 
similar setups. Such lattices are created by the intersection of 
laser fields, and they are nondissipative periodic potential energy 
surfaces for the atoms. Motivated by this success in experimentally 
simulating the single-species BH model with ultracold atomic Bose gases
loaded into optical lattices, there has been recently an intense 
theoretical activity in analyzing BH as well as Fermi-Hubbard 
type models~\cite{bloch08}.

For instance, in addition to the Mott insulator and single-species
superfluid phases, it has been predicted that the two-species 
BH model has at least two additional phases: an incompressible 
super-counter flow and a compressible paired superfluid 
phase~\cite{kuklov03, altman03, kuklov04, isacsson05, arguelles07, trefzger09, anzi09, buonsante09, hubener09, menotti10}. 
Our main interest here is in the latter phase, where a direct 
transition from the Mott insulator to the paired superfluid phase 
(superfluidity of composite bosons, i.e. Bose-Bose pairs) has been
predicted, when both species have integer fillings and the 
interspecies interaction is sufficiently large and attractive.
Given that the interspecies interactions can be fine tuned in ongoing 
experiments, e.g. with $^{41}$K-$^{87}$Rb with 
mixtures~\cite{catani08, thalhammer08}, via using Feshbach 
resonances, we hope that some of our results could be tested with 
ultracold atomic systems.

In this paper, we examine the ground-state phase diagram of the 
two-species BH model with on-site boson-boson interactions in 
$d$-dimensional hypercubic lattices, including both the repulsive 
and attractive interspecies interaction, via a strong-coupling 
perturbation theory in the hopping. We carry the expansion out 
to third-order in the hopping, and perform a scaling analysis 
using the known critical behavior at the tip of the insulating lobes, 
which allows us to accurately predict the critical point, and the shape 
of the insulating lobes in the plane of the chemical potential 
and the hopping. This technique was previously used to discuss 
the phase diagram of the single-species BH 
model~\cite{freericks96, kuhner98, buonsante04, sengupta05, freericks09}, extended BH 
model~\cite{iskin09}, and of the hardcore BH model with a 
superlattice~\cite{hen10}, and its results showed an excellent 
agreement with Monte Carlo simulations~\cite{freericks09, hen10}. 
Motivated by the success of this technique with these models, here 
we apply it to the two-species BH model, hoping to develop 
an analytical approach which could be as accurate as the numerical ones. 

The remaining paper is organized as follows. After introducing 
the model Hamiltonian in Sec.~\ref{sec:tsbh}, we develop the strong-coupling 
expansion in Sec.~\ref{sec:sce}, where we derive an analytical 
expression for the phase boundary between the incompressible Mott 
insulator and the compressible superfluid phase. Then, in Sec.~\ref{sec:ext}, 
we propose a chemical-potential extrapolation technique based 
on scaling theory to extrapolate our third-order power-series 
expansion into a functional form that is appropriate for the Mott 
lobes, and use it to obtain typical ground-state phase diagrams.
A brief summary of our conclusions is given in Sec.~\ref{sec:conclusions}.

\section{Two-species Bose-Hubbard Model}
\label{sec:tsbh}

To describe Bose-Bose mixtures loaded into optical lattices, we consider 
the following two-species BH Hamiltonian,
\begin{align}
\label{eqn:ebhh}
H = &- \sum_{i,j, \sigma} t_{ij,\sigma} b_{i,\sigma}^\dagger b_{j,\sigma} 
  + \sum_{i,\sigma} \frac{U_{\sigma\sigma}}{2} \widehat{n}_{i,\sigma} (\widehat{n}_{i,\sigma}-1) \nonumber \\
&+ U_{\uparrow\downarrow}\sum_{i} \widehat{n}_{i,\uparrow} \widehat{n}_{i,\downarrow} - \sum_{i,\sigma} \mu_\sigma \widehat{n}_{i,\sigma},
\end{align}
where the pseudo-spin $\sigma \equiv \{\uparrow, \downarrow\}$ labels the 
trapped hyperfine states of a given species of bosons, or labels different 
types of bosons in a two-species mixture, $t_{ij, \sigma}$ is the 
tunneling (or hopping) matrix between sites $i$ and $j$, 
$b_{i,\sigma}^\dagger$ ($b_{i,\sigma}$) is the boson creation (annihilation) 
and $\widehat{n}_{i,\sigma} = b_{i,\sigma}^\dagger b_{i,\sigma}$ 
is the boson number operator at site $i$, $U_{\sigma\sigma'}$ is the strength 
of the onsite boson-boson interaction between $\sigma$ and $\sigma'$ 
components, and $\mu_\sigma$ is the chemical potential.
In this manuscript, we consider a $d$-dimensional hypercubic lattice with $M$ sites,
for which we assume $t_{ij,\sigma}$ is a real symmetric matrix with elements
$t_{ij,\sigma} = t_\sigma \ge 0$ for $i$ and $j$ nearest neighbors and $0$ otherwise.
The lattice coordination number (or the number of nearest neighbors) for such
lattices is $z = 2d$.

We take the intraspecies interactions to be repulsive ($\{U_{\uparrow\uparrow}, 
U_{\downarrow\downarrow}\} > 0$), but discuss both repulsive and attractive 
interspecies interaction $U_{\uparrow\downarrow}$ as long as 
$U_{\uparrow\uparrow} U_{\downarrow\downarrow} > U_{\uparrow\downarrow}^2$.
This guarantees the stability of the mixture against collapse when
$U_{\uparrow\downarrow} \ll 0$, and against phase separation when 
$U_{\uparrow\downarrow} \gg 0$. However, when the interspecies interaction is 
sufficiently large and attractive, we note that instead of a direct transition 
from the Mott insulator to a single particle superfluid phase, 
it is possible to have a transition from the Mott insulator to a paired 
superfluid phase (superfluidity of composite bosons, i.e. Bose-Bose 
pairs)~\cite{kuklov03, altman03, kuklov04, isacsson05, arguelles07, trefzger09, anzi09, buonsante09, hubener09, menotti10}. 
Therefore, one needs to consider both possibilities, as discussed next.

\section{Strong-Coupling Expansion}
\label{sec:sce}

We use the many-body version of Rayleigh-Schr\"odinger perturbation 
theory in the kinetic energy term to perform the expansion (in powers 
of $t_\uparrow$ and $t_\downarrow$) for the different energies needed 
to carry out our analysis. The strong-coupling expansion technique was 
previously used to discuss the phase diagram of the single-species
BH model~\cite{freericks96, kuhner98, buonsante04, freericks09}, extended BH 
model~\cite{iskin09}, and of the hardcore BH model with a 
superlattice~\cite{hen10}, and its results showed an excellent agreement 
with Monte Carlo simulations~\cite{freericks09, hen10}. 
Motivated by the success of this technique with these models, here we 
apply it to the two-species BH model.

To determine the phase boundary separating the incompressible Mott phase 
from the compressible superfluid phase within the strong-coupling expansion 
method, one needs the energy of the Mott phase and of its `defect' states 
(those states which have exactly one extra elementary particle or hole about 
the ground state) as a function of $t_\uparrow$ and $t_\downarrow$. 
At the point where the energy of the incompressible state becomes equal 
to its defect state, the system becomes compressible, assuming that 
the compressibility approaches zero continuously at the phase boundary. 
Here, we remark that this technique cannot be used to calculate the 
phase boundary between two compressible phases.

\subsection{Ground-State Wave Functions}
\label{sec:wf}

The perturbation theory is performed with respect to the ground state of 
the system when $t_\uparrow = t_\downarrow = 0$, and therefore we first 
need zeroth order wave functions of the Mott phase and of its defect states.
To zeroth order in $t_\uparrow$ and $t_\downarrow$, the Mott insulator 
wave function can be written as,
\begin{eqnarray}
\label{eqn:wf-ins}
|\Psi_{\rm Mott}^{\rm ins (0)} \rangle &=& 
\frac{1}{\sqrt{n_\uparrow ! n_\downarrow !}} \prod_{i}
(b_{i,\uparrow}^\dagger)^{n_\uparrow} (b_{i,\downarrow}^\dagger)^{n_\downarrow} | 0 \rangle,
\end{eqnarray}
where $\langle \widehat{n}_{i,\sigma} \rangle = n_\sigma$ is an integer number
corresponding to the ground-state occupancy of the pseudo-spin $\sigma$ bosons,
$\langle \cdots \rangle$ is the thermal average, and $| 0 \rangle$ is the vacuum state. 
On the other hand, the wave functions of the defect states are determined 
by degenerate perturbation theory. The reason for that lies in the 
fact that when exactly one extra elementary particle or hole is added 
to the Mott phase, it could go to any of the $M$ lattice sites, 
since all of those states share the same energy when $t_\uparrow = t_\downarrow = 0$. 
Therefore, the initial degeneracy of the defect states is of order $M$.

When the elementary excitations involve a single-$\sigma$-particle (exactly 
one extra pseudo-spin $\sigma$ boson) or a single-$\sigma$-hole (exactly 
one less pseudo-spin $\sigma$ boson), this degeneracy is lifted at first 
order in $t_\uparrow$ and $t_\downarrow$. The treatment for this case 
is very similar to the single-species BH model~\cite{freericks96, iskin09}, 
and the wave functions (to zeroth order in $t_\uparrow$ and $t_\downarrow$) 
for the single-$\sigma$-particle and single-$\sigma$-hole defect states 
turn out to be
\begin{eqnarray}
\label{eqn:wf-sph}
|\Psi_{\rm def}^{\rm s \sigma p (0)} \rangle &=& \frac{1}{\sqrt{n_\sigma+1}}
\sum_{i} f_i^{\rm s \sigma p} b_{i,\sigma}^\dagger |\Psi_{\rm Mott}^{\rm ins (0)} \rangle, \\
|\Psi_{\rm def}^{\rm s \sigma h (0)} \rangle &=& \frac{1}{\sqrt{n_\sigma}}
\sum_{i} f_i^{\rm s \sigma h} b_{i,\sigma} |\Psi_{\rm Mott}^{\rm ins (0)} \rangle,
\end{eqnarray}
where $f_i^{\rm s \sigma p} = f_i^{\rm s \sigma h}$ is the eigenvector 
of the hopping matrix $t_{ij,\sigma}$ with the highest eigenvalue 
(which is $z t_\sigma$ with $z = 2d$) such that
$
\sum_{j} t_{ij,\sigma} f_j^{\rm s \sigma p} = z t_\sigma f_i^{\rm s \sigma p}.
$
The normalization condition requires that $\sum_{i} |f_i^{\rm s \sigma p}|^2 = 1$.
Notice that we choose the highest eigenvalue of $t_{ij,\sigma}$ because the 
hopping matrix enters the Hamiltonian as $-t_{ij,\sigma}$, and we ultimately 
want the lowest-energy states. 

However, when the elementary excitations involve two particles (exactly one 
extra boson of each species) or two holes (exactly one less boson of each species),
the degeneracy is lifted at second order in $t_\uparrow$ and 
$t_\downarrow$. Such elementary excitations occur when $U_{\uparrow\downarrow}$ 
is sufficiently large and attractive~\cite{sufficient}, 
and the wave functions (to zeroth order in $t_\uparrow$ and $t_\downarrow$) 
for the two-particle and two-hole defect states can be written as
\begin{eqnarray}
\label{eqn:wf-tph}
|\Psi_{\rm def}^{\rm tp (0)} \rangle &=& \frac{1}{\sqrt{(n_\uparrow+1)(n_\downarrow+1)}}
\sum_{i} f_i^{\rm tp} b_{i,\uparrow}^\dagger b_{i,\downarrow}^\dagger |\Psi_{\rm Mott}^{\rm ins (0)} \rangle, \\
|\Psi_{\rm def}^{\rm th (0)} \rangle &=& \frac{1}{\sqrt{n_\uparrow n_\downarrow}}
\sum_{i} f_i^{\rm th} b_{i,\uparrow} b_{i,\downarrow} |\Psi_{\rm Mott}^{\rm ins (0)} \rangle,
\end{eqnarray}
where $f_i^{\rm tp} = f_i^{\rm th}$ turns out to be the eigenvector 
of the $t_{ij,\uparrow} t_{ij,\downarrow}$ matrix with the highest 
eigenvalue (which is $z t_\uparrow t_\downarrow$ with $z = 2d$) 
such that
$
\sum_{j} t_{ij,\uparrow} t_{ij,\downarrow} f_j^{\rm tp} = z t_\uparrow t_\downarrow f_i^{\rm tp}.
$
Since the elementary excitations involve two particles or two holes, the 
degenerate defect states cannot be connected by one hopping, but rather 
require two hoppings to be connected. Therefore, one expects the 
degeneracy to be lifted at least at second order in $t_\uparrow$ 
and $t_\downarrow$, as discussed next.

\subsection{Ground-State Energies}
\label{sec:en}

Next, we employ the many-body version of Rayleigh-Schr\"odinger perturbation 
theory in $t_\uparrow$ and $t_\downarrow$ with respect to the ground state of 
the system when $t_\uparrow = t_\downarrow = 0$, and calculate the energy 
of the Mott phase and of its defect states. The energy of the Mott state 
is obtained via nondegenerate perturbation theory, and to third order 
in $t_\uparrow$ and $t_\downarrow$ it is given by
\begin{align}
\label{eqn:en-mott}
\frac{E_{\rm Mott}^{\rm ins}}{M} &= \sum_{\sigma} \frac{U_{\sigma\sigma}}{2} n_\sigma (n_\sigma-1) 
+ U_{\uparrow\downarrow} n_\uparrow n_\downarrow - \sum_\sigma \mu_\sigma n_\sigma  \nonumber \\ 
&- \sum_{\sigma} n_\sigma (n_\sigma+1) \frac{z t_\sigma^2}{U_{\sigma\sigma}} + O(t^4).
\end{align}
This is an extensive quantity, i.e. $E_{\rm Mott}^{\rm ins}$ is proportional 
to the number of lattice sites $M$. The odd-order terms in $t_\uparrow$ and 
$t_\downarrow$ vanish for the $d$-dimensional hypercubic lattices considered 
in this manuscript, which is simply because the Mott state given in 
Eq.~(\ref{eqn:wf-ins}) cannot be connected to itself by only one hopping, 
but rather requires two hoppings to be connected.
Notice that Eq.~(\ref{eqn:en-mott}) recovers the known result for the 
single-species BH model when one of the pseudo-spin components have 
vanishing filling, e.g. $n_{\downarrow} = 0$~\cite{freericks96, iskin09}.

The calculation of the defect-state energies is more involved since it requires 
using degenerate perturbation theory. As mentioned above, when the elementary 
excitations involve a single-$\sigma$-particle or a single-$\sigma$-hole, 
the degeneracy is lifted at first order in $t_\uparrow$ and $t_\downarrow$.
A lengthy but straightforward calculation leads to the energy of the 
single-$\sigma$-particle defect state up to third order in $t_\uparrow$ 
and $t_\downarrow$ as
\begin{widetext}
\begin{align}
\label{eqn:en-spar}
E_{\rm def}^{\rm s \sigma p} &= E_{\rm Mott}^{\rm ins} + U_{\uparrow\downarrow} n_{-\sigma} 
  + U_{\sigma\sigma} n_\sigma - \mu_\sigma - (n_\sigma+1) z t_\sigma \nonumber \\
&- n_\sigma \left[ \frac{n_\sigma+2}{2} + (n_\sigma+1)(z-3) \right] \frac{z t_\sigma^2}{U_{\sigma\sigma}}
  -2n_{-\sigma}(n_{-\sigma}+1) \frac{U_{\uparrow\downarrow}^2}{U_{-\sigma-\sigma}^2-U_{\uparrow\downarrow}^2} 
  	\frac{z t_{-\sigma}^2}{U_{-\sigma-\sigma}} \nonumber \\
&- n_\sigma (n_\sigma+1) \left[ n_\sigma(z-1)^2 + (n_\sigma+1)(z-1)(z-4) + (n_\sigma+2)(3z/4-1) 
  \right] \frac{z t_\sigma^3}{U_{\sigma\sigma}^2} \nonumber \\
&-4(n_{\sigma}+1)n_{-\sigma}(n_{-\sigma}+1)
  \frac{U_{\uparrow\downarrow}^2}{U_{-\sigma-\sigma}^2-U_{\uparrow\downarrow}^2}
  \left( z-1 - \frac{U_{-\sigma-\sigma}^2}{U_{-\sigma-\sigma}^2-U_{\uparrow\downarrow}^2} \right)
  \frac{z t_\sigma t_{-\sigma}^2}{U_{-\sigma-\sigma}^2}
  + O(t^4),
\end{align}
where $(-\uparrow) \equiv \downarrow$ and vice versa. 
Here, we assume $U_{\sigma\sigma} \gg t_\sigma$ and 
$\{U_{-\sigma-\sigma}, |U_{-\sigma-\sigma} \pm U_{\uparrow\downarrow}| \} \gg t_{-\sigma}$.
Equation~(\ref{eqn:en-spar}) is valid for all $d$-dimensional hypercubic lattices, and it
recovers the known result for the single species BH model when 
$n_{-\sigma} = 0$~\cite{freericks96, iskin09}. Note that this expression
also recovers the known result for the single species BH model when 
$U_{\uparrow\downarrow} = 0$, which provides an independent check of the algebra.
To third order in $t_\uparrow$ and $t_\downarrow$, we obtain a similar expression 
for the energy of the single-$\sigma$-hole defect state given by
\begin{align}
\label{eqn:en-shol}
E_{\rm def}^{\rm s \sigma h} &= E_{\rm Mott}^{\rm ins} - U_{\uparrow\downarrow} n_{-\sigma} 
  - U_{\sigma\sigma} (n_\sigma-1) + \mu_\sigma - n_\sigma z t_\sigma \nonumber \\
&- (n_\sigma+1) \left[ \frac{n_\sigma-1}{2} + n_\sigma(z-3) \right] \frac{z t_\sigma^2}{U_{\sigma\sigma}}
  -2n_{-\sigma}(n_{-\sigma}+1) \frac{U_{\uparrow\downarrow}^2}{U_{-\sigma-\sigma}^2-U_{\uparrow\downarrow}^2} 
  	\frac{z t_{-\sigma}^2}{U_{-\sigma-\sigma}} \nonumber \\
&- n_\sigma (n_\sigma+1) \left[ (n_\sigma+1)(z-1)^2 + n_\sigma(z-1)(z-4) + (n_\sigma-1)(3z/4-1) 
  \right] \frac{z t_\sigma^3}{U_{\sigma\sigma}^2} \nonumber \\
&-4n_{\sigma}n_{-\sigma}(n_{-\sigma}+1)
  \frac{U_{\uparrow\downarrow}^2}{U_{-\sigma-\sigma}^2-U_{\uparrow\downarrow}^2}
  \left( z-1 - \frac{U_{-\sigma-\sigma}^2}{U_{-\sigma-\sigma}^2-U_{\uparrow\downarrow}^2} \right)
  \frac{z t_\sigma t_{-\sigma}^2}{U_{-\sigma-\sigma}^2}
  + O(t^4),
\end{align}
which is also valid for all $d$-dimensional hypercubic lattices, and it also
recovers the known result for the single-species BH model 
when $n_{-\sigma} = 0$ or $U_{\uparrow\downarrow} = 0$~\cite{freericks96, iskin09}.
Here, we again assume $U_{\sigma\sigma} \gg t_\sigma$ and 
$\{U_{-\sigma-\sigma}, |U_{-\sigma-\sigma} \pm U_{\uparrow\downarrow}| \} \gg t_{-\sigma}$.
We also checked the accuracy of the second-order terms in 
Eqs.~(\ref{eqn:en-spar}) and~(\ref{eqn:en-shol}) via exact small-cluster 
(two-site) calculations with one $\sigma$ and two $-\sigma$ particles.

We note that the mean-field phase boundary between the Mott phase and its 
single-$\sigma$-particle and single-$\sigma$-hole defect states 
can be calculated as
\begin{align}
\label{eqn:mf-mu}
\mu_\sigma^{\rm par, hol} &= U_{\sigma\sigma}(n_\sigma-1/2) + U_{\uparrow\downarrow} n_{-\sigma} - z t_\sigma/2 
	\pm \sqrt{U_{\sigma\sigma}^2/4 - U_{\sigma\sigma}(n_\sigma+1/2)z t_\sigma + z^2 t_\sigma^2/4}.
\end{align}
This expression is exact for infinite-dimensional hypercubic 
lattices, and it recovers the known result for the single species BH model 
when $n_{-\sigma} = 0$ or $U_{\uparrow\downarrow} = 0$~\cite{fisher}. 
In the $d \to \infty$ limit (while keeping $dt_{\sigma}$ constant), we 
checked that our strong-coupling perturbation results given in 
Eqs.~(\ref{eqn:en-spar}) and~(\ref{eqn:en-shol}) agree with this 
exact solution when the latter is expanded out to third order in $t_\uparrow$ 
and $t_\downarrow$, providing an independent check of the algebra.
Equation~(\ref{eqn:mf-mu}) also shows that, for infinite-dimensional lattices, 
the Mott lobes are separated by $U_{\uparrow\downarrow} n_{-\sigma}$, but 
their shapes and critical points (the latter are obtained by setting 
$\mu_\sigma^{\rm par} = \mu_\sigma^{\rm hol}$) are independent of $U_{\uparrow\downarrow}$.
This is not the case for finite-dimensional lattices as can be clearly
seen from our results. It is also important to mention here that both the 
shapes and critical points are independent of the sign of 
$U_{\uparrow\downarrow}$ in finite dimensions (at the third-order presented 
here) as can be seen in Eqs.~(\ref{eqn:en-spar}) and~(\ref{eqn:en-shol}).

However, when the elementary excitations involve two particles or two holes (which
occurs when $U_{\uparrow\downarrow}$ is sufficiently large and attractive~\cite{sufficient}),
the degeneracy is lifted at second order in $t_\uparrow$ and $t_\downarrow$. 
A lengthy but straightforward calculation leads to the energy of the two-particle 
defect state up to third order in $t_\uparrow$ and $t_\downarrow$ as
\begin{align}
\label{eqn:en-tpar}
E_{\rm def}^{\rm tp} &= E_{\rm Mott}^{\rm ins} + U_{\uparrow\downarrow} (n_\uparrow+n_\downarrow+1)
+ \sum_\sigma \left( U_{\sigma\sigma} n_\sigma - \mu_\sigma \right)
+ \frac{2(n_\uparrow+1)(n_\downarrow+1)}{U_{\uparrow\downarrow}} z t_\uparrow t_\downarrow \nonumber \\
&+ \sum_\sigma
	\left[ \frac{(n_\sigma+1)^2}{U_{\uparrow\downarrow}} 
	- \frac{n_\sigma(n_\sigma+2)}{2U_{\sigma\sigma}+U_{\uparrow\downarrow}} 
	+ \frac{2n_\sigma(n_\sigma+1)}{U_{\sigma\sigma}} 
	\right] z t_\sigma^2 + O(t^4).
\end{align}
Here, we assume 
$\{ U_{\sigma\sigma}, |U_{\uparrow\downarrow}|, 2U_{\sigma\sigma}+U_{\uparrow\downarrow} \} \gg t_\sigma$.
Equation~(\ref{eqn:en-tpar}) is valid for all $d$-dimensional hypercubic lattices, 
where the odd-order terms in $t_\uparrow$ and $t_\downarrow$ vanish~\cite{infinite}.
To third order in $t_\uparrow$ and $t_\downarrow$, we obtain a similar 
expression for the energy of the two-hole defect state given by
\begin{align}
\label{eqn:en-thol}
E_{\rm def}^{\rm th} &= E_{\rm Mott}^{\rm ins} - U_{\uparrow\downarrow} (n_\uparrow+n_\downarrow-1)
- \sum_\sigma \left[ U_{\sigma\sigma} (n_\sigma-1) - \mu_\sigma \right]
+ \frac{2n_\uparrow n_\downarrow}{U_{\uparrow\downarrow}} z t_\uparrow t_\downarrow \nonumber \\
&+ \sum_\sigma
	\left[ \frac{n_\sigma^2}{U_{\uparrow\downarrow}} 
	- \frac{(n_\sigma^2-1)}{2U_{\sigma\sigma}+U_{\uparrow\downarrow}} 
	+ \frac{2n_\sigma(n_\sigma+1)}{U_{\sigma\sigma}} 
	\right] z t_\sigma^2 + O(t^4),
\end{align}
\end{widetext}
which is also valid for all $d$-dimensional hypercubic lattices, where the 
odd-order terms in $t_\uparrow$ and $t_\downarrow$ vanish~\cite{infinite}.
Here, we again assume 
$\{ U_{\sigma\sigma}, |U_{\uparrow\downarrow}|, 2U_{\sigma\sigma}+U_{\uparrow\downarrow} \} \gg t_\sigma$.
Since the single-$\sigma$-particle and single-$\sigma$-hole 
defect states have corrections to first order in the hopping, while 
the two-particle and two-hole defect states have corrections to second 
order in the hopping, the slopes of the Mott lobes are finite 
as $\{t_\uparrow, t_\downarrow\} \rightarrow 0$ in the former case, 
but they vanish in the latter case. Hence, the shape of the insulating
lobes are expected to be very different for two-particle or two-hole
excitations. In addition, the chemical-potential widths ($\mu_\sigma$) 
of all Mott lobes are $U_{\sigma\sigma}$ in the former case,
but they [$(\mu_\uparrow+\mu_\downarrow)/2$] are 
$U_{\uparrow\downarrow} + (U_{\uparrow\uparrow} + U_{\downarrow\downarrow})/2$
in the latter.

We note that in the limit when $t_\uparrow = t_\downarrow = t$, 
$U_{\uparrow\uparrow} = U_{\downarrow\downarrow} = U_0$, 
$U_{\uparrow\downarrow} = U'$, $n_\uparrow = n_\downarrow = n_0$, 
$\mu_\uparrow = \mu_\downarrow = \mu$, and $z = 2$ (or $d = 1$), 
Eq.~(\ref{eqn:en-thol}) is in complete agreement with Eq.~(3) of
Ref.~\cite{arguelles07}, providing an independent check of the algebra.
In addition, in the limit when $t_\uparrow = t_\downarrow = J$, 
$U_{\uparrow\uparrow} = U_{\downarrow\downarrow} = U$,
$U_{\uparrow\downarrow} = W \approx -U$, $n_\uparrow = n_\downarrow = m$, 
and $\mu_\uparrow = \mu_\downarrow = \mu$, Eqs.~(\ref{eqn:en-tpar}) 
and~(\ref{eqn:en-thol}) reduce to those given in Ref.~\cite{trefzger09}
(after setting $U_{NN} = 0$ there). However, the terms that are 
proportional to $t_\uparrow t_\downarrow$ are not included in their 
definitions of the two-particle and two-hole excitation gaps. 
We also checked the accuracy of Eqs.~(\ref{eqn:en-tpar}) and~(\ref{eqn:en-thol})
via exact small-cluster (two-site) calculations with one particle
of each species.

We would also like to remark in passing that the energy difference between
the Mott phase and its defect states determine the phase boundary of 
the particle and hole branches. This is because at the point where 
the energy of the incompressible state becomes equal to its defect state, 
the system becomes compressible, assuming that the compressibility 
approaches zero continuously at the phase boundary. While $E_{\rm Mott}^{\rm ins}$ 
and its defects $E_{\rm def}^{\rm s \sigma p}$, $E_{\rm def}^{\rm s \sigma h}$, 
$E_{\rm def}^{\rm tp}$ and $E_{\rm def}^{\rm th}$ depend on the lattice 
size $M$, their difference do not. Therefore, the chemical potentials 
that determine the particle and hole branches are independent of 
$M$ at the phase boundaries. This indicates that the numerical 
Monte Carlo simulations should not have a strong dependence on $M$.

It is known that the third-order strong-coupling expansion is not very 
accurate near the tip of the Mott lobes, as $t_\uparrow$ and $t_\downarrow$ 
are not very small there~\cite{freericks96, iskin09}. For this reason, 
an extrapolation technique is highly desirable to determine more accurate 
phase diagrams. Therefore, having discussed the third-order strong-coupling 
expansion for a general two-species Bose-Bose mixtures with 
arbitary hoppings $t_\sigma$, interactions $U_{\sigma\sigma'}$, 
densities $n_\sigma$, and chemical potentials $\mu_\sigma$, next we 
show how to develop a scaling theory.

\section{Extrapolation Technique}
\label{sec:ext}

In this section, we propose a chemical potential extrapolation technique 
based on scaling theory to extrapolate our third-order power-series 
expansion into a functional form that is appropriate for the entire Mott 
lobes. It is known that the critical point at the tip of the lobes
has the scaling behavior of a ($d+1$)-dimensional $XY$ model, and therefore 
the lobes have Kosterlitz-Thouless shapes for $d = 1$ and power-law shapes 
for $d > 1$. For illustration purposes, here we analyze only the latter case, 
but this technique can be easily adapted to the $d = 1$ case~\cite{freericks96}.

\subsection{Scaling Ansatz}
\label{sec:ansatz}

From now on we consider a two-species mixture with 
$t_\uparrow = t_\downarrow = t$, $U_{\uparrow\uparrow} = 
U_{\downarrow\downarrow} = U$, $U_{\uparrow\downarrow} = V$, $n_\uparrow 
= n_\downarrow = n$, and $\mu_\uparrow = \mu_\downarrow = \mu$.
When $d > 1$, we propose the following ansatz which includes the 
known power-law critical behavior of the tip of the lobes
\begin{align}
\label{eqn:smu}
\frac{\mu^{\pm}}{U} = A(x) \pm B(x)(x_c-x)^{z\nu},
\end{align}
where 
$
A(x) = a + b x + c x^2 + d x^3 + \cdots
$ 
and 
$
B(x) = \alpha + \beta x + \gamma x^2 + \delta x^3 + \cdots
$
are regular functions of $x = 2dt/U$, $x_c$ is the critical point which 
determines the location of the lobes, and $z\nu$ is the critical 
exponent for the ($d+1$)-dimensional $XY$ model which determines the 
shape of the lobes near $x_c = 2dt_c/U$. In Eq.~(\ref{eqn:smu}), the plus sign 
corresponds to the particle branch, and the minus sign corresponds to 
the hole branch. The form of the ansatz is taken to be the same for both 
single- and two-partice (or single- and two-hole) excitations, 
but the parameters are very different.

The parameters $a$, $b$, $c$ and $d$ depend on $U$, $V$ and $n$, and 
they are determined by matching them with the coefficients given by our 
third-order expansion such that
$
A(x) = (\mu^{\rm par} + \mu^{\rm hol})/(2U).
$
Here, $\mu^{\rm par}$ and $\mu^{\rm hol}$ are our strong-coupling expansion
results determined from Eqs.~(\ref{eqn:en-spar}) and~(\ref{eqn:en-shol})
for the single-particle and single-hole excitations, or from 
Eqs.~(\ref{eqn:en-tpar}) and~(\ref{eqn:en-thol}) for the two-particle and
two-hole excitations, respectively.
Writing our strong-coupling expansion results for the particle and hole 
branches in the form $\mu^{\rm par} = U\sum_{n = 0}^3 e_n^{+} x^n$ and 
$\mu^{\rm hol} = U\sum_{n = 0}^3 e_n^{-} x^n$, leads to 
$a = (e_0^+ + e_0^-)/2$, $b = (e_1^+ + e_1^-)/2$, $c = (e_2^+ + e_2^-)/2$,
and $d = (e_3^+ + e_3^-)/2$. To determine the $U$, $V$ and $n$ 
dependence of the parameters $\alpha$, $\beta$, $\gamma$, $\delta$, 
$x_c$ and $z\nu$, we first expand the left hand side of 
$
B(x)(x_c-x)^{z\nu} = (\mu^{\rm par} - \mu^{\rm hol})/(2U)
$
in powers of $x$, and match the coefficients with the coefficients given by our
third-order expansion, leading to
\begin{align}
\alpha &= \frac{e_0^+-e_0^-}{2x_c^{z\nu}}, 
\label{eqn:alpha} \\
\frac{\beta}{\alpha} &= \frac{z\nu}{x_c} 
	+ \frac{e_1^+-e_1^-}{e_0^+-e_0^-}, 
\label{eqn:beta} \\
\frac{\gamma}{\alpha} &= \frac{z\nu(z\nu+1)}{2x_c^2} 
	+ \frac{z\nu}{x_c} \frac{e_1^+-e_1^-}{e_0^+-e_0^-} 
	+ \frac{e_2^+-e_2^-}{e_0^+-e_0^-}, 
\label{eqn:gamma} \\
\frac{\delta}{\alpha} &= \frac{z\nu(z\nu+1)(z\nu+2)}{6x_c^3} 
	+ \frac{z\nu(z\nu+1)}{2x_c^2} \frac{e_1^+-e_1^-}{e_0^+-e_0^-} \nonumber \\
& + \frac{z\nu}{x_c} \frac{e_2^+-e_2^-}{e_0^+-e_0^-} + \frac{e_3^+-e_3^-}{e_0^+-e_0^-}.
\label{eqn:delta} 
\end{align}
We fix $z\nu$ at its well-known values such that $z\nu \approx 2/3$ for 
$d = 2$ and $z\nu = 1/2$ for $d > 2$. If the exact value of $x_c$ is 
known via other means, e.g. numerical simulations, $\alpha$, $\beta$, 
$\gamma$ and $\delta$ can be calculated accordingly, for which the 
extrapolation technique gives very accurate results~\cite{freericks09, hen10}.
If the exact value of $x_c$ is not known, then we set $\delta = 0$, 
and solve Eqs.~(\ref{eqn:alpha}),~(\ref{eqn:beta}),~(\ref{eqn:gamma})
and the $\delta = 0$ equation to determine 
$\alpha$, $\beta$, $\gamma$ and $x_c$ self-consistently, 
which also leads to accurate results~\cite{freericks96, iskin09}.
Next we present typical ground-state phase diagrams for $(d=2)$- and 
($d = 3$)-dimensional hypercubic lattices obtained from this 
extrapolation technique.

\begin{figure} [htb]
\centerline{\scalebox{0.6}{\includegraphics{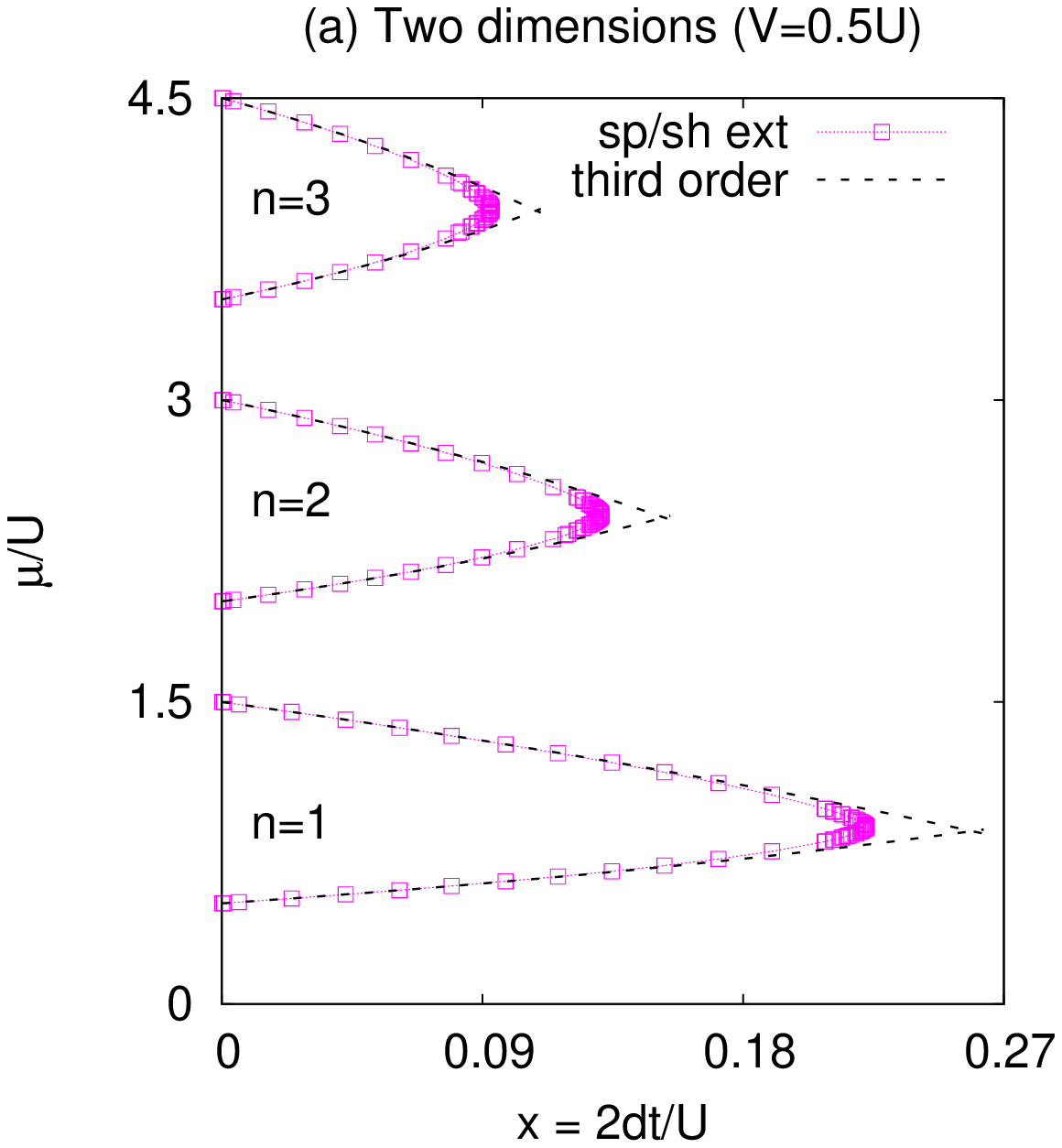}}}
\centerline{\scalebox{0.6}{\includegraphics{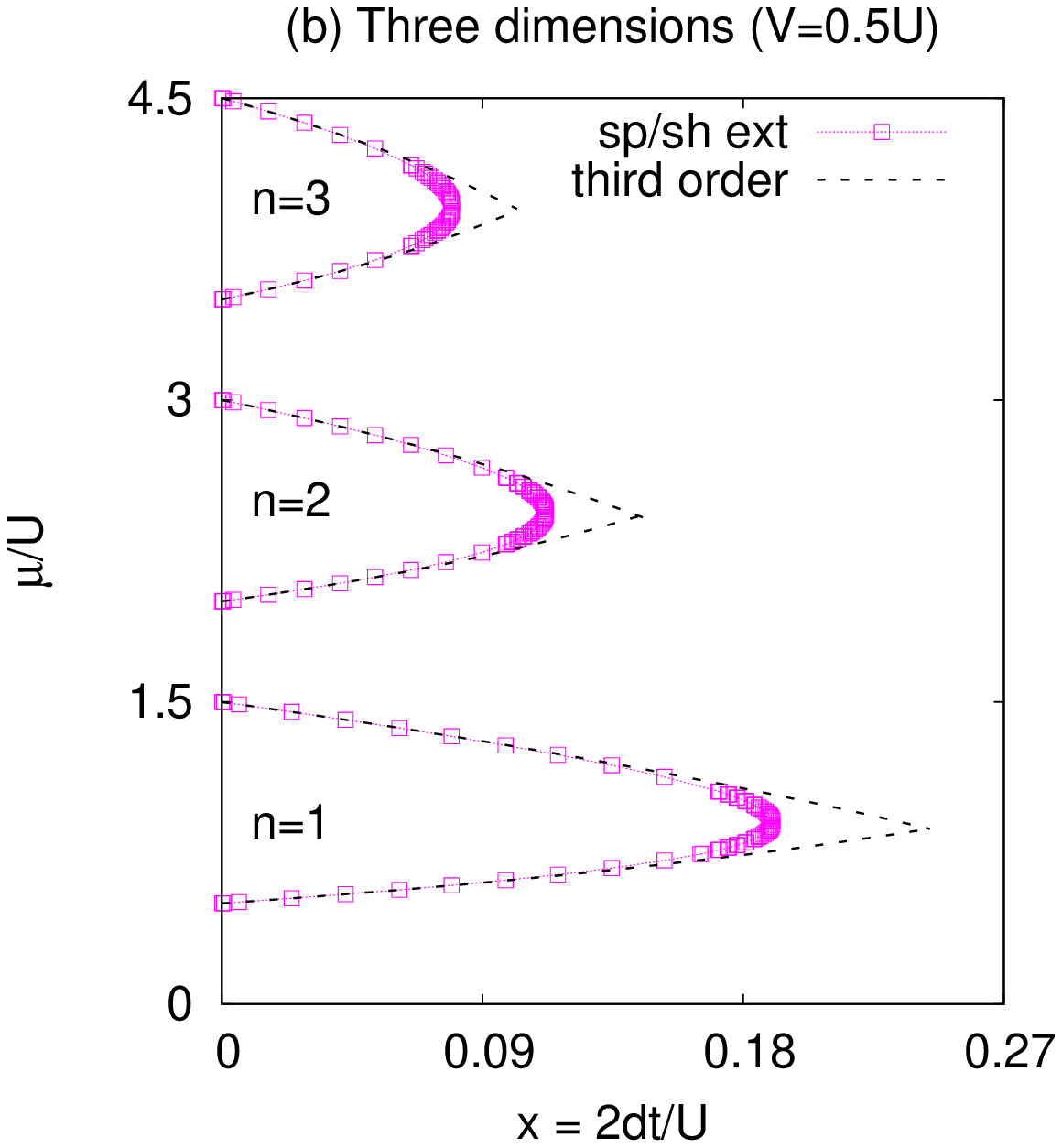}}}
\caption{\label{fig:single} (Color online)
Chemical potential $\mu$ (in units of $U$) versus $x = 2dt/U$ 
phase diagram for (a) two- and (b) three-dimensional 
hypercubic lattices with $t_\uparrow = t_\downarrow = t$, 
$U_{\uparrow\uparrow} = U_{\downarrow\downarrow} = U$, 
$U_{\uparrow\downarrow} = V = 0.5U$, $n_\uparrow = n_\downarrow = n$, 
and $\mu_\uparrow = \mu_\downarrow = \mu$. 
The dotted lines correspond to phase boundary for the Mott 
insulator to superfluid state as determined from the third-order 
strong-coupling expansion, and the hollow pink-squares
to the extrapolation fit for the single-particle or single-hole 
excitations discussed in the text. Recall that an incompressible 
super-counter flow phase also exists outside of the Mott insulator lobes.
}
\end{figure}
\subsection{Numerical Results}
\label{sec:numerics}

\begin{figure} [htb]
\centerline{\scalebox{0.6}{\includegraphics{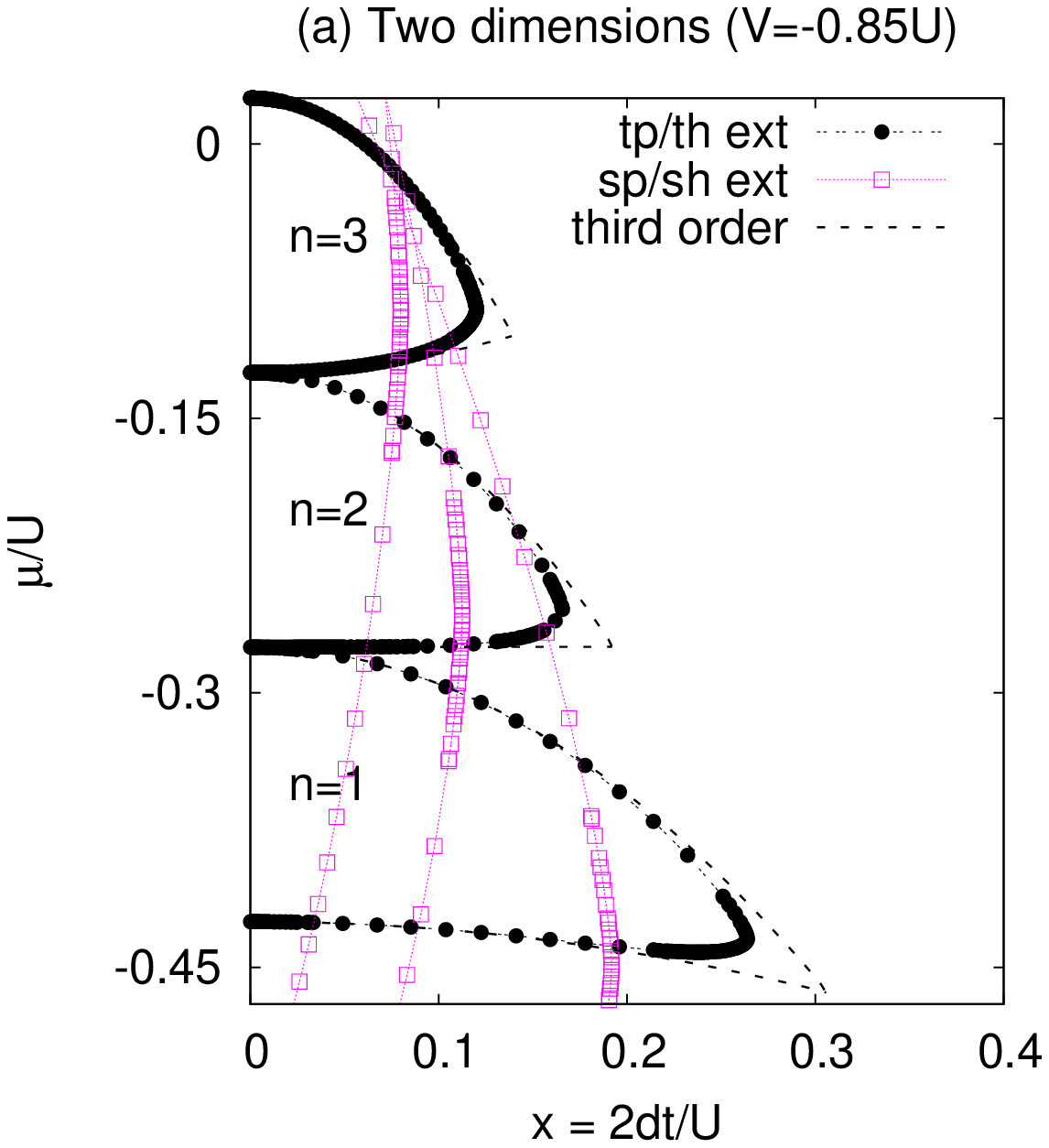}}}
\centerline{\scalebox{0.6}{\includegraphics{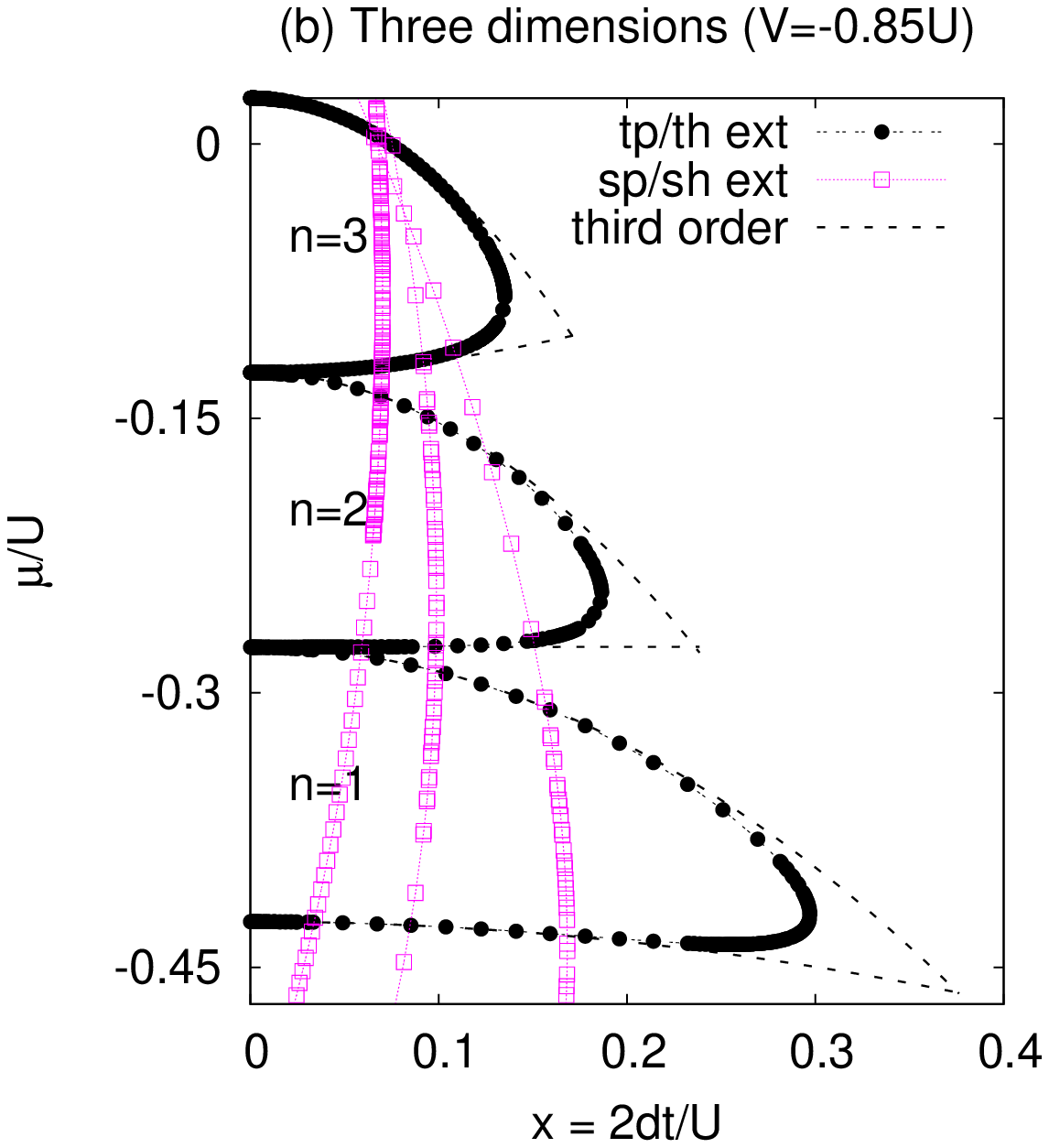}}}
\caption{\label{fig:pair} (Color online)
Chemical potential $\mu$ (in units of $U$) versus $x = 2dt/U$ 
phase diagram for (a) two- and (b) three-dimensional 
hypercubic lattices with $t_\uparrow = t_\downarrow = t$, 
$U_{\uparrow\uparrow} = U_{\downarrow\downarrow} = U$, 
$U_{\uparrow\downarrow} = V = -0.85U$, $n_\uparrow = n_\downarrow = n$, 
and $\mu_\uparrow = \mu_\downarrow = \mu$. 
The dotted lines correspond to phase boundary for the Mott 
insulator to superfluid state determined from the third-order 
strong-coupling expansion, the hollow pink-squares
to the extrapolation fit for the single-particle or single-hole 
excitations (shown only for illustration purposes), and the solid 
black-circles to the extrapolation fit for the 
two-particle or two-hole excitations (the 
expected transition) discussed in the text.
}
\end{figure}

In Figs.~\ref{fig:single} and~\ref{fig:pair}, the results of the 
third-order strong-coupling expansion (dotted lines) are compared 
to those of the extrapolation technique (hollow pink-squares and
solid black-circles) when $V = 0.5U$ and $V = -0.85U$, respectively,
in two ($d = 2$ or $z = 4$) and three ($d = 3$ or $z = 6$) dimensions.
We recall here that $t_\uparrow = t_\downarrow = t$, $U_{\uparrow\uparrow} = 
U_{\downarrow\downarrow} = U$, $U_{\uparrow\downarrow} = V$, 
$n_\uparrow = n_\downarrow = n$, and $\mu_\uparrow = \mu_\downarrow = \mu$.

In Fig.~\ref{fig:single}, we show the chemical potential $\mu$ 
(in units of $U$) versus $x = 2dt/U$ phase diagram for (a) two-dimensional 
and (b) three-dimensional hypercubic lattices, where
we choose the interspecies interaction to be repulsive $V = 0.5U$.
Comparing Eqs.~(\ref{eqn:en-spar}) and~(\ref{eqn:en-shol})
with Eqs.~(\ref{eqn:en-tpar}) and~(\ref{eqn:en-thol}), we expect that 
the excited state of the system to be the usual superfluid for 
all $V > 0$ for all $t$. 
The dotted lines correspond to phase boundary for the Mott 
insulator to superfluid state as determined from the third-order 
strong-coupling expansion, and the hollow pink-squares correspond 
to the extrapolation fits for the single-particle and single-hole 
excitations discussed in the text. 
We recall here that an incompressible super-counter flow 
phase~\cite{kuklov03, altman03, kuklov04, anzi09} also exists 
outside of the Mott insulator lobes, but our current formalism cannot 
be used to locate its phase boundary.

\begin{center}
\begin{table} [htb]
\caption{\label{table:repulsive} 
List of the critical points (location of the tips) $x_c = 2dt_c/U$ 
for the first two Mott insulator lobes that are found from the 
chemical potential extrapolation technique described in the text.
Here, $t_\uparrow = t_\downarrow = t$, $U_{\uparrow\uparrow} 
= U_{\downarrow\downarrow} = U$, $U_{\uparrow\downarrow} = V$, 
$n_\uparrow = n_\downarrow = n$, and $\mu_\uparrow = \mu_\downarrow = \mu$.
These critical points for the single-particle or single-hole
excitations are determined from Eqs.~(\ref{eqn:en-spar}) 
and~(\ref{eqn:en-shol}), and they tend to move in as $V$ increases,
and are independent of the sign of $V$.
}
\begin{tabular}{c|cc|cc}
\hline \hline
& $d = 2$ & & $d = 3$ & \\
\cline{2-5}
$V/U$ & $n = 1$ & $n = 2$ & $n = 1$ & $n = 2$ \\
\hline
0.0  & 0.234	& 0.138	& 0.196	& 0.116	 \\
0.1  & 0.234	& 0.138	& 0.196	& 0.115	 \\
0.2  & 0.233	& 0.137	& 0.195	& 0.115	 \\
0.3  & 0.230	& 0.136	& 0.194	& 0.114	 \\
0.4  & 0.227	& 0.134	& 0.193	& 0.113	 \\
0.5  & 0.223	& 0.131	& 0.190	& 0.112	 \\
0.6  & 0.217	& 0.128	& 0.187	& 0.110	 \\
0.7  & 0.208	& 0.123	& 0.182	& 0.107	 \\
0.8  & 0.197	& 0.116	& 0.174	& 0.102	 \\
0.9  & 0.193	& 0.113	& 0.163	& 0.095	 \\
\hline \hline
\end{tabular}
\end{table}
\end{center}

At $t = 0$, the chemical potential width of all Mott lobes are $U$
(similar to the single-species BH model), but they are separated 
from each other by $V$ as a function of $\mu$. 
As $t$ increases from zero, the range of $\mu$ about which the ground state 
is a Mott insulator decreases, and the Mott insulator phase disappears 
at a critical value of $t$, beyond which the system becomes a superfluid.
In addition, similar to what was found for the single-species 
BH model~\cite{freericks96, iskin09}, the strong-coupling expansion 
overestimates the phase boundaries, and it leads to unphysical pointed 
tips for all Mott lobes, which is expected since a finite-order 
expansion cannot describe the physics of the critical point correctly.
A short list of $V/U$ versus the critical points $x_c = 2dt_c/U$ 
is presented for the first two Mott insulator lobes in 
Table~\ref{table:repulsive}, where it is shown that the critical 
points tend to move in as $V$ increases. This is because presence 
of a second species (say $-\sigma$ ones) screens the onsite 
intraspecies repulsion $U_{\sigma\sigma}$ between $\sigma$-species, 
and hence increases the superfluid region.

In Fig.~\ref{fig:pair}, we show the chemical potential $\mu$ 
(in units of $U$) versus $x = 2dt/U$ phase diagram for (a) two-dimensional 
and (b) three-dimensional hypercubic lattices, where in these 
figures we choose the interspecies interaction to be attractive 
$V = -0.85U$. Comparing Eqs.~(\ref{eqn:en-spar}) and~(\ref{eqn:en-shol})
with Eqs.~(\ref{eqn:en-tpar}) and~(\ref{eqn:en-thol}), we expect that 
the excited state of the system to be a paired superfluid for all 
$V < 0$ when $t \to 0$. This is clearly seen in the figure where
the dotted lines correspond to phase boundary for the Mott 
insulator to superfluid state as determined from the third-order 
strong-coupling expansion, the hollow pink-squares correspond 
to the extrapolation fits for the single-particle and single-hole 
excitations (shown only for illustration purposes), 
and the solid black-circles correspond to the extrapolation 
fits for the two-particle and two-hole excitations (this is the 
expected transition) discussed in the text. 

At $t = 0$, the chemical potential width of all Mott lobes are 
$V+U = 0.15U$, which is in contrast with the single-species BH model. 
As $t$ increases from zero, the range of $\mu$ about which the ground state 
is a Mott insulator decreases here as well, and the Mott insulator 
phase disappears at a critical value of $t$, beyond which the system 
becomes a paired superfluid. The strong-coupling expansion 
again overestimates the phase boundaries, and it again leads 
to unphysical pointed tips for all Mott lobes.
In addition, a short list of $V/U$ versus the critical points 
$x_c = 2dt_c/U$ are presented for the first two Mott insulator lobes in 
Table~\ref{table:repulsive}. Our results are consistent with the 
expectation that, for small $V$, the locations of the tips increase 
as a function of $V$, because the presence of a nonzero $V$ is what 
allowed these states to form in the first place. However, when 
$V$ is larger than some critical value ($\sim 0.6U$), the locations of 
the tips decrease, and they eventually vanish when $V =-U$. This may 
indicate an instability towards a collapse since at this point
$U_{\uparrow\uparrow} U_{\downarrow\downarrow}$ is exactly equal to 
$U_{\uparrow\downarrow}^2$.

\begin{center}
\begin{table} [htb]
\caption{\label{table:attractive} 
List of the critical points (location of the tips) $x_c = 2dt_c/U$ 
that are found from the chemical potential extrapolation technique 
described in the text. 
Here, $t_\uparrow = t_\downarrow = t$, $U_{\uparrow\uparrow} 
= U_{\downarrow\downarrow} = U$, $U_{\uparrow\downarrow} = V$, 
$n_\uparrow = n_\downarrow = n$, and $\mu_\uparrow = \mu_\downarrow = \mu$. 
These critical points for the two-particle or two-hole excitations 
are determined from Eqs.~(\ref{eqn:en-tpar}) and~(\ref{eqn:en-thol})
when $V < 0$. Note that, for small $V$, $x_c$'s tend to increase as 
a function of $V$, since the presence of a nonzero $V$ is what allowed 
these states to form in the first place. However, $x_c$'s decrease
beyond a critical $V$, and they eventually vanish when $V =-U$, 
which may indicate an instability towards a collapse.
}
\begin{tabular}{c|cc|cc}
\hline \hline
& $d = 2$ & & $d = 3$ & \\
\cline{2-5}
$V/U$ & $n = 1$ & $n = 2$ & $n = 1$ & $n = 2$ \\
\hline
-0.01  & 0.0543		&	0.0337	&	0.0611	&	0.0379	 \\
-0.03  & 0.0937		&	0.0582	&	0.105		&	0.0655	 \\
-0.05  & 0.121		&	0.0749	&	0.136		&	0.0843	 \\
-0.07  & 0.142		&	0.0883	&	0.160		&	0.0994	 \\
-0.1	 & 0.169		&	0.105		&	0.190		&	0.118		 \\
-0.2	 & 0.233		&	0.145		&	0.262		&	0.164		 \\
-0.3	 & 0.277		&	0.173		&	0.311		&	0.195		 \\
-0.4	 & 0.307		&	0.193		&	0.345		&	0.217		 \\
-0.5	 & 0.325		&	0.205		&	0.366		&	0.230		 \\
-0.6	 & 0.331		&	0.209		&	0.372		&	0.235		 \\
-0.7	 & 0.321		&	0.203		&	0.362		&	0.228		 \\
-0.8	 & 0.291		&	0.183		&	0.327		&	0.206		 \\
-0.9	 & 0.225		&	0.141		&	0.253		&	0.159		 \\
-0.93	 & 0.193		&	0.121	&	0.217	&	0.136	 \\
-0.95	 & 0.166		&	0.103	&	0.187	&	0.116	 \\
-0.97	 & 0.1304		&	0.0812	&	0.147	&	0.0913	 \\
-0.99	 & 0.0764		&	0.0474	&	0.0860	&	0.0534	 \\
\hline \hline
\end{tabular}
\end{table}
\end{center}

Compared to the $V > 0$ case shown in Fig.~\ref{fig:single}, 
note that shape of the Mott insulator to paired superfluid phase 
boundary is very different, showing a re-entrant behavior in 
all dimensions from paired superfluid to Mott insulator and 
again to a paired superfluid phase, as a function of $t$. Our results
are consistent with an early numerical time-evolving block decimation 
(TEBD) calculation~\cite{arguelles07}, where such a re-entrant quantum phase 
transition in one dimension was predicted. 

The re-entrant quantum phase transition occurs when coefficient 
of the hopping term in Eq.~(\ref{eqn:en-thol}) is negative 
[so that the two-hole excitation branch has a negative slope 
in $(\mu_\uparrow + \mu_\downarrow)/2$ versus $t_\sigma$ phase diagram 
when $t_\sigma \to 0$], i.e.
$
-(2n_\uparrow n_\downarrow/U_{\uparrow\downarrow}) z t_\uparrow t_\downarrow
-\sum_\sigma[ n_\sigma^2/U_{\uparrow\downarrow} 
	- (n_\sigma^2-1)/(2U_{\sigma\sigma}+U_{\uparrow\downarrow}) 
	+ 2n_\sigma(n_\sigma+1)/U_{\sigma\sigma}] z t_\sigma^2
$
term, which occurs for the first few Mott lobes beyond a critical 
$U_{\uparrow\downarrow}$. When this coefficient is negative, its
value is most negative for the first Mott lobe, and therefore 
the effect is strongest there. However, the coefficient increases 
and eventually becomes positive as a function of filling, and thus 
the re-entrant behavior becomes weaker as filling increases, and 
it eventually disappears beyond a critical filling. For the 
parameters used in Fig.~\ref{fig:pair}, this occurs only for 
the first lobe, as can be seen in the figures.
We also note that the sign of this coefficient is independent of 
the dimensionality of the lattice, since $z = 2d$ enters into 
the coefficient only as an overall factor. 

What happens when $t_\uparrow \ne t_\downarrow$ and/or $U_{\uparrow\uparrow} \ne 
U_{\downarrow\downarrow}$?
We donot expect any qualitative change for attractive interspecies 
interactions. However, for repulsive interspecies interactions, this 
lifts the degeneracy of the single-particle or single-hole excitation energies. 
While the transition is from a double Mott insulator to a double 
superfluid of both species in the degenerate case, it is from a double-Mott 
insulator of both species to a Mott insulator of one species and a 
superfluid of the other in the nondegenerate case.

\section{Conclusions}
\label{sec:conclusions}

We analyzed the zero temperature phase diagram of the two-species 
Bose-Hubbard (BH) model with on-site boson-boson interactions in 
$d$-dimensional hypercubic lattices, including both the repulsive 
and attractive interspecies interaction. We used the many-body 
version of Rayleigh-Schr\"odinger perturbation theory in the 
kinetic energy term with respect to the ground state of the system 
when the kinetic energy term is absent, and calculate ground state 
energies needed to carry out our analysis. This technique was 
previously used to discuss the phase diagram of the single-species
BH model~\cite{freericks96, kuhner98, buonsante04, freericks09}, extended BH 
model~\cite{iskin09}, and of the hardcore BH model with a 
superlattice~\cite{hen10}, and its results showed an excellent agreement 
with Monte Carlo simulations~\cite{freericks09, hen10}. 
Motivated by the success of this technique with these models, here 
we generalized it to the two-species BH model, hoping to develop 
an analytical approach which could be as accurate as the numerical ones. 

We derived analytical expressions for the phase boundary between the 
incompressible Mott insulator and the compressible superfluid phase 
up to third order in the hoppings. We also proposed a chemical 
potential extrapolation technique based on the scaling theory 
to extrapolate our third-order power series expansion into a 
functional form that is appropriate for the Mott lobes. In particular,
when the interspecies interaction is sufficiently large and attractive, 
we found a re-entrant quantum phase transition from paired superfluid 
(superfluidity of composite bosons, i.e. Bose-Bose pairs) to Mott 
insulator and again to a paired superfluid in all one, two and three 
dimensions. 
Since the available Monte Carlo calculations~\cite{kuklov04, isacsson05} 
do not provide the Mott insulator to superfluid transition phase boundary 
in the experimentally more relevant chemical potential versus hopping 
plane, we could not compare our results with them. This comparison 
is highly desirable to judge the accuracy of our strong-coupling 
expansion results.

A possible direction to extend this work is to consider the limit 
where hopping of one-species is much larger than the other. 
In this limit, the two-species BH model reduces to the Bose-Bose version 
of the Falicov-Kimball model~\cite{falicov69}, the Fermi-Fermi 
version of which has been widely discussed in the condensed-matter
literature and the Fermi-Bose version has just been studied~\cite{iskin09dmft}.  
It is known for such models that there is a tendency towards both phase 
separation and density wave order~\cite{soyler09}, which requires a 
new calculation partially similar to that of Ref.~\cite{iskin09}.
One can also examine how the momentum distribution changes with the 
hopping in the insulating phases~\cite{freericks09, iskin09md},
which has direct relevance to ultracold atomic experiments.

\section{Acknowledgments}
\label{sec:ack}

The author thanks Anzi Hu, L. Mathey and J. K. Freericks for discussions, 
and The Scientific and Technological Research Council of Turkey 
(T\"{U}B$\dot{\mathrm{I}}$TAK) for financial support.


\begin{thebibliography}{99}

\bibitem{fisher} 
M. P. A. Fisher, P. B. Weichman, G. Grinstein, and D. S. Fisher, 
Phys. Rev. B \textbf{40}, 546 (1989).

\bibitem{greiner02} 
M. Greiner, O. Mandel, T. Esslinger, T. W. H\"ansch, and I. Bloch,
Nature (London) \textbf{415}, 39 (2002). 

\bibitem{stoferle04} 
T. St\"oferle, H. Moritz, C. Schori, M. K\"ohl, and T. Esslinger,
Phys. Rev. Lett. \textbf{92}, 130403 (2004).

\bibitem{spielman07} 
I. B. Spielman, W. D. Phillips, and J. V. Porto, 
Phys. Rev. Lett. \textbf{98}, 080404 (2007).

\bibitem{spielman08} 
I. B. Spielman, W. D. Phillips, and J. V. Porto, 
Phys. Rev. Lett. \textbf{100}, 120402 (2008).

\bibitem{bloch08} 
I. Bloch, J. Dalibard, and W. Zwerger,
Rev. Mod. Phys. \textbf{80}, 885 (2008).

\bibitem{kuklov03}
A. B. Kuklov and B. V. Svistunov,
Phys. Rev. Lett. \textbf{90}, 100401 (2003).

\bibitem{altman03}
E. Altman, W. Hofstetter, E. Demler, and M. D. Lukin,
New J. Phys. \textbf{5}, 113 (2003).

\bibitem{kuklov04}
A. Kuklov, N. Prokof'ev, and B. Svistunov, 
Phys. Rev. Lett. \textbf{92}, 050402 (2004).

\bibitem{isacsson05}
A. Isacsson, Min-Chul Cha, K. Sengupta, and S. M. Girvin, 
Phys. Rev. B \textbf{72}, 184507 (2005).

\bibitem{arguelles07} 
A. Arg\"uelles and L. Santos, 
Phys. Rev. A \textbf{75}, 053613 (2007).

\bibitem{trefzger09} 
C. Trefzger, C. Menotti, and M. Lewenstein,
Phys. Rev. Lett. \textbf{103}, 035304 (2009).

\bibitem{anzi09} 
Anzi Hu, L. Mathey, I. Danshita, E. Tiesinga, C. J. Williams, and C. W. Clark,
Phys. Rev. A \textbf{80}, 023619 (2009).

\bibitem{buonsante09}
P. Buonsante, S. M. Giampaolo, F. Illuminati, V. Penna, and A. Vezzani, 
Eur. Phys. J. B \textbf{68}, 427 (2009).

\bibitem{hubener09}
A. Hubener, M. Snoek, and W. Hofstetter,
Phys. Rev. B \textbf{80}, 245109 (2009).

\bibitem{menotti10}
C. Menotti and S. Stringari,
Phys. Rev. A \textbf{81}, 045604 (2010).

\bibitem{catani08}
J. Catani, L. De Sarlo, G. Barontini, F. Minardi, and M. Inguscio,
Phys. Rev. A \textbf{77}, 011603(R) (2008).

\bibitem{thalhammer08}
G. Thalhammer, G. Barontini, L. De Sarlo, J. Catani, F. Minardi, and M. Inguscio,
Phys. Rev. Lett. \textbf{100}, 210402 (2008).

\bibitem{freericks96} 
J. K. Freericks and H. Monien, 
Phys. Rev. B \textbf{53}, 2691 (1996).

\bibitem{kuhner98}
T. D. K\"uhner and H. Monien, 
Phys. Rev. B, \textbf{58}, R14741 (1998).

\bibitem{buonsante04}
P. Buonsante, V. Penna, and A. Vezzani, 
Phys. Rev. B \textbf{70}, 184520 (2004).

\bibitem{sengupta05}
K. Sengupta and N. Dupuis, 
Phys. Rev. A \textbf{71}, 033629 (2005).

\bibitem{freericks09} 
J. K. Freericks, H. R. Krishnamurthy, Y. Kato, N. Kawashima, and N. Trivedi, 
Phys. Rev. A \textbf{79}, 053631 (2009).

\bibitem{iskin09}
M. Iskin and J. K. Freericks, 
Phys. Rev. A \textbf{79}, 053634 (2009).

\bibitem{hen10}
Itay Hen, M. Iskin, and M. Rigol, 
Phys. Rev. B \textbf{81}, 064503 (2010).

\bibitem{sufficient}
Recall that $U_{\uparrow\downarrow}^2$ cannot be greater than or
equal to $U_{\uparrow\uparrow} U_{\downarrow\downarrow}$, 
otherwise the mixture would be unstable against collapse. In addition,
see e.g. Fig.~7 in~\cite{anzi09}, where TEBD calculations show in one 
dimension that $V \lesssim -0.06U$ is already sufficient for the Mott
insulator to paired superfluid transition.

\bibitem{infinite}
Note that, unlike those of single-particle and single-hole excitations
where $d t_\sigma$ is a constant when $d \to \infty$, 
in the case of two-particle and two-hole excitations, 
$d t_\sigma^2$ must be kept constant when $d \to \infty$. 
In this respect, Eqs.~(\ref{eqn:en-tpar}) and~(\ref{eqn:en-thol}) 
do not contain any finite-$d$ correction at the second order in hopping.

\bibitem{falicov69} 
L. M. Falicov and J. C. Kimball, 
Phys. Rev. Lett. \textbf{22}, 997 (1969).

\bibitem{iskin09dmft}
M. Iskin and J. K. Freericks, 
Phys. Rev. A \textbf{80}, 053623 (2009); and see references therein.

\bibitem{soyler09}
\c{S}. G. S\"oyler, B. Capogrosso-Sansone, N. V. Prokof'ev, and B. V. Svistunov,
New J. Phys. \textbf{11}, 073036 (2009).

\bibitem{iskin09md}
M. Iskin and J. K. Freericks, 
Phys. Rev. A \textbf{80}, 063610 (2009).

\end{thebibliography}
\end{document}